\def\ra{\rangle}
\def\la{\langle}
\def\t{\tilde}

\documentclass[aps,amsmath,amssymb,amsfonts,showpace,twocolumn,pra]{revtex4}
\usepackage{amscd}
\usepackage{amsthm}
\usepackage{graphicx}

\def\qed{\leavevmode\unskip\penalty9999 \hbox{}\nobreak\hfill
     \quad\hbox{\leavevmode  \hbox to.77778em{%
               \hfil\vrule   \vbox to.675em%
               {\hrule width.6em\vfil\hrule}\vrule\hfil}}
     \par\vskip3pt}

\begin{document}
\title{Unextendible maximally entangled bases in $\mathbb{C}^{d}\bigotimes\mathbb{C}^{d'}$}
\author{Mao-Sheng Li,Yan-Ling Wang,Zhu-Jun Zheng}

\affiliation{{Department of Mathematics,
 South China University of Technology, Guangzhou
510640,China}\\
}

\begin{abstract}
We solved the unextendible maximally
entangled basis (UMEB) problem in $\mathbb{C}^{d}\bigotimes\mathbb{C}^{d'}(d\neq d')$,the results turn
out to be that there always exist a UMEB.In addition,there might be two sets of UMEB with different
numbers.The main difficult is to prove the unextendibility of the set of states.We give an explicit
 construction of UMEB by considering the Schmidt number of the complementary space of
the states we construct.

\end{abstract}

\maketitle

\section{Introduction}

Entanglement plays important role in quantum information, such as teleportation, quantum error correction and
quantum secret sharing\cite{EofCon,nils,mhor}.The unextendible product basis (UPB) has been extensively investigated.
Considerable elegant results have been obtained with interesting applications to
the theory of quantum information \cite{d03,UPB1}. The UPB is a set of incomplete orthonormal
product basis whose complementary space has no product states.
It was also shown that the space complementary to a UPB contains bound entanglement \cite{Be}.Moreover,
the states comprising a UPB are not distinguishable by local measurements and classical communication.

There it was shown that there are sets of orthogonal product vectors of a tensor
 product Hilbert space $\mathbb{C}^{d}\bigotimes\mathbb{C}^{d'}(d\neq d')$  such that there are no
 further product states orthogonal to every state in the set, even though the size of the set is smaller than $dd'$.

Recently S. Bravyi and J. A.
Smolin generalized the notion of the UPB to unextendible maximally
entangled basis\cite {s3}:
a set of orthonormal maximally entangled states in
$\mathbb{C}^{d}\bigotimes\mathbb{C}^{d}$ consisting of fewer than $d^2$ vectors which have no additional maximally
entangled vectors orthogonal to all of them.
In \cite {s3},the authors proved that there does not exist UMEBs for $d=2$ and constructed
 a 6-member UMEB for $d=3$ and a 12-member UMEB for $d=4$.And the authors left us a question:whether the  nonsquare UMEBS
 exist or not? In addition,there are some authors study the locally unextendible non-maximally entangled basis
(LUNMEB)\cite{IP}.

In \cite{BC}, the authors studied the  UMEB in  $\mathbb{C}^{d}\bigotimes\mathbb{C}^{d'}$ ($\frac{d'}{2}<d<d'$).
They constructed a
$d^{2}$-member UMEB and also left a quetion:whether there exist UMEB in other case $\frac{d'}{2}\geq d$ or not.

In this paper,we study the UMEB in arbitrary bipartite spaces.We give an explicit construction of UMEBs in
$\mathbb{C}^{d}\bigotimes\mathbb{C}^{d'} (d<d')$,hence we state that the UMEB  exists if $d<d'$ and this gives an
answer to the question asked in \cite{s3,BC}.We can give explicit expression of the vectors in the completementary
space of the constructed states.Then we can state that there is no maximally entangled states in the completementary
space of the constructed states by considering their Schmidt number.

\bigskip

\textbf{Definition\cite{BC}.} A set of states
\{$|\phi_{i}\rangle\in\mathbb{C}^{d}\bigotimes\mathbb{C}^{d'}:i=1,2,\cdots,n,\,n<dd'$\}
is called an $n$ member UMEB if and only if

(i) $|\phi_{i}\rangle$, $i=1,2,\cdots,n$, are maximally entangled;

(ii) $\langle\phi_{i}|\phi_{j}\rangle=\delta_{ij}$;

(iii) if $\langle\phi_{i}|\psi\rangle=0$ for all $i=1,2,\cdots,n$, then $|\psi\rangle$ cannot be maximally entangled.

Here state $|\psi\rangle$ is said to be a $d\otimes d'$ maximally entangled state if and only if
for arbitrary given orthonormal complete basis \{$|i_{A}\rangle$\}
of subsystem A, there exists an orthonormal basis
\{$|i_{B}\rangle$\} of subsystem B such that $|\psi\rangle$ can be written as
$|\psi\rangle=\frac{1}{\sqrt{d}}\sum_{i=0}^{d-1}|i_{A}\rangle\otimes|i_{B}\rangle$ \cite{s4}.

\section{UMEBs in $\mathbb{C}^{d}\bigotimes\mathbb{C}^{d'}\,(d'>d)$}
A $d^2$ member UMEB in $\mathbb{C}^{d}\bigotimes\mathbb{C}^{d'}(\frac{d'}{2}<d<d')$ has been constructed in \cite{BC}
as the following.
$$|\phi_{mn}\rangle={\frac{1}{\sqrt{d}}}{\sum_{p=0}^{d-1}}\zeta_d^{np}|p\oplus m\rangle |p'\rangle$$
where $\zeta_{d}=e^{\frac{2\pi\sqrt{-1}}{d}}$, $k\oplus m$ denotes
$(k+m)$ mod $d$

\bigskip

If we take a look at the  condition $\frac{d'}{2}<d<d'$ mention above, we can find that this condition is just
equivalent with $ d'=d+r, 0<r<d$.

\bigskip

It's no wonder that  the problem of UMEB in $\mathbb{C}^{d}\bigotimes\mathbb{C}^{d'}$ with ${d'}=dq+r,\ \ {0<r<d}$ is a
 generalized  problem.

\bigskip
\noindent{\bf Proposition  \ 1.}
In $\mathbb{C}^{d}\bigotimes\mathbb{C}^{d'} ,{d'}=dq+r,{0<r<d}$,for any  integer
$$ m,n=0,1\cdots,d-1;l=1,\cdots,q,$$ we define $$|\phi_{nml}\rangle\triangleq
{\frac{1}{\sqrt{d}}}{\sum_{p=0}^{d-1}}\zeta_d^{np}|p\oplus m\rangle |((l-1)d+p)'\rangle $$
where  $k\oplus m$ denotes $ (k+m)$ mod $d$.\\

Then these $qd^2$ states are  unextendible mutually orthogonal maximally entangled states.

\bigskip
Proof:
(i) Orthogonality\\
\\
$\la\phi_{\t{n} \t{m} \t{l}}|\phi_{nml}\ra$\\
{
\small
$$=\frac{1}{d}\sum_{p=0}^{d-1}\sum_{\t{p}=0}^{d-1}\zeta_{d}^{np-{\t{n}\t{p}}}
\langle((\t{l}-1)d+\t{p})'|\langle\t{p}\oplus\t{m}|p\oplus m\rangle|((l-1)d+p)'\rangle$$
$$=\frac{1}{d}\sum_{p=0}^{d-1}\sum_{\t{p}=0}^{d-1}\zeta_{d}^{np-{\t{n}\t{p}}}
\la\t{p}\oplus\t{m}|p\oplus m\ra\la((\t{l}-1)d+\t{p})'|((l-1)d+p)'\ra$$
}
\noindent$=\frac{1}{d}\sum_{p=0}^{d-1}\zeta_{d}^{(n-\t{n})p}
\la p\oplus\t{m}|p\oplus m\ra \delta_{l\t{l}}$\\
\\
$=\delta_{m\t{m}}\delta_{n\t{n}}\delta_{l\t{l}}$\\

\bigskip

(ii)Maximally entangled.\\

This can be easily checked by the definition of $|\phi_{mnl}\ra$.

\bigskip

(iii)Unextendible \\

Let $V_1$ denotes the subspace span by $$\{ |\phi_{nml}\ra,n,m=0,1,\cdots,d-1;l=1,\cdots,q\}.$$
We have
$$Dim(V_1)=qd^2,$$
so $$Dim({V_1}^\perp)=dd'-qd^2=dr.$$

Now let $$|\psi_{i,j}\ra=|ij\ra,i=0,1,\cdots,d-1;j=qd,\cdots,qd+r-1$$

Let $V_2$ denotes the subspace span by $$\{ |\psi_{i,j}\ra,i=0,1,\cdots,d-1;j=qd,\cdots,qd+r-1\}.$$

Following easily calculation,we have
$$\la\psi_{i^{'},j^{'}}|\phi_{i,j}\ra=0$$

Because $Dim(V_2)=dr$,so $V_1^\perp=V_2.$
$\forall v\in V_2$  has the bellowing  form

$$v=\sum_{i=0}^{d-1}\sum_{\j=0}^{r-1}a_{i,j}|i\ra|qd+j\ra.$$
So the Schmidt rank of every vector in $V_2$  is no higher than $r$ which is less than $d$,hence we derived
that any state  in $V_1^\perp$ is not maximally entangled.

From (i)(ii)(iii) and $qd^2<dd'$,we can conclude that the  states $\{|\phi_{i,j}\ra\}$ consist of a $qd^2$ member UMEB
 in $\mathbb{C}^{d}\bigotimes\mathbb{C}^{d'}.$    \qed

\bigskip

In \cite{BC}, the authors ask whether the set of $d^2$ member orthonormal maximally entagled states they constructed is
unextendible or not in the case of $d\leq\frac{d'}{2}$.Proposition 1 give a deny answer to this question
when $d'=qd+r,q>1,0<r<d$.

\bigskip
\noindent{\bf Example 1.}We find an 8 member UMEB in $\mathbb{C}^{2}\bigotimes\mathbb{C}^{5}$,this example
is not satisfied the condition in \cite{BC}.Each row represent a states but not normal for the purpose of the consise
notation.
{\begin{table}[h]
 \caption{8-member UMEB in $\mathbb{C}^{2}\bigotimes\mathbb{C}^{5}$}

$\begin{array}{cccccccccc}\hline\hline
             00 & 01 & 10 & 11 & 02 & 03 & 12 & 13 & 04 & 14 \\ \hline
             1 & 0 & 0 & 1 & 0 & 0 & 0 & 0 & 0 & 0 \\
             1 & 0 & 0 & -1 & 0 & 0 & 0 & 0 & 0 & 0 \\
             0 & 1 & 1 & 0 & 0 & 0 & 0 & 0 & 0 & 0 \\
             0 & 1 & -1 & 0 & 0 & 0 & 0 & 0 & 0 & 0 \\
             0 & 0 & 0 & 0 & 1 & 0 & 0 & 1 & 0 & 0 \\
             0 & 0 & 0 & 0 & 1 & 0 & 0 & -1 & 0 & 0 \\
             0 & 0 & 0 & 0 & 0 & 1 & 1 & 0 & 0 & 0 \\
             0 & 0 & 0 & 0 & 0 & 1 & -1 & 0 & 0 & 0 \\ \hline\hline
           \end{array}$

\end{table}

From the table above,we can observe that the four states can be look as the Bell states in
$\mathbb{C}^{2}\bigotimes\mathbb{C}^{2}$ with the base $|00\ra,|01\ra,|10\ra,|11\ra$ and the last
four states also can be seen as the Bell states with the base $|02\ra,|03\ra,|12\ra,|13\ra.$

\bigskip

Now we notice that the above construction can not be efficient if $r=0$,for $qd^2=dd' $ in this case.So we give another
construction to find the UMEB for other cases.A nature question arise when we consider the UMEB problem.
Are there two sets of UMEB with different number? The Proposition below gives a positive answer to this problem
when $(d\geq3,d'-d\geq2)$.

\bigskip
\noindent{\bf Proposition  \ 2.} In $\mathbb{C}^{d}\bigotimes\mathbb{C}^{d'}(d<d')$,for any integer

{\footnotesize
$$m\in\left\{
             \begin{array}{ll}
          \{d'-1,d'-2,...,d'-d+1\} & d'\geq 2d \\
           \{d'-1,d'-2,...,d'-r\}  & d<d'<2d, d'=d+r
             \end{array}
           \right.$$
}

let
$${|\phi_{i,j}\ra}={\frac{1}{\sqrt{d}}}{\sum_{k=0}^{d-1}}\ e^{\frac{2\pi \sqrt{-1}}{d}ki}|k\rangle |k\oplus j\rangle,$$
$$ i=0,...,d-1;j=0,...,m-1.$$
where $k\oplus j$ means $k+j$ mod m.\\

Then $\{|\phi_{i,j}\ra\}$ is a $dm$ member UMEB.

\bigskip
Proof:(i) Orthongonality

  $\la \phi_{i^{'},j^{'}}|\phi_{i,j}\ra$\\
  $=\frac{1}{d}{\sum_{k=0}^{d-1}}{\sum_{k^{'}=0}^{d-1}}\ e^{\frac{2\pi \sqrt{-1}}{d}(ki-k^{'}i^{'})}
    \la k^{'}|k \ra \la k{'}\oplus j{'}|k\oplus j\ra$\\
  \\
  $=\frac{1}{d}{\sum_{k=0}^{d-1}}\ e^{\frac{2\pi \sqrt{-1}}{d}k(i-i^{'})}\la k{'}\oplus j{'}|k\oplus j\ra$\\
  \\
  $=\frac{1}{d}\delta_{j,j^{'}}{\sum_{k=0}^{d-1}}\ e^{\frac{2\pi \sqrt{-1}}{d}k(i-i^{'})}$\\
  \\
  $=\delta_{j,j^{'}}\delta_{i,i^{'}}$\\

  (ii)Maximally entangled

  By the definition

  $${|\phi_{i,j}\ra}={\frac{1}{\sqrt{d}}}{\sum_{k=0}^{d-1}}\ e^{\frac{2\pi \sqrt{-1}}{d}ki}|k\rangle |k\oplus j\rangle,\ $$ $$i=0,...,d-1;j=0,...,m-1.$$

  $k\oplus j$ is just $k+j$ mod m,because $m>d$,so for $0\leq k,k'\leq d-1$, $k\oplus j=k'\oplus j$ if and only with $k=k'$.
So ${|\phi_{i,j}\ra}$ are maximally entangled states.

\bigskip
(iii)Unextendible

Similar with the proof in Proposition 1,we denote
$$V_1=span\{|\phi_{i,j}\ra,i=0,...,d-1;j=0,...,m-1\}$$
$$V_2=span\{|\psi_{i,j}\ra=|ij\ra,i=0,...,d-1;j=m,...,d'-1\}$$
It's clear that
$$Dim(V_1)=dm,Dim(V_2)=d(d'-m).$$

By $\la \psi_{i',j'}|\phi_{i,j}\ra=0$,we derived that $V_1^\perp=V_2$.Now every vector $v\in V_2$ has the following
form

$$v=\sum_{i=0}^{d-1}\sum_{j=0}^{d'-m-1}a_{i,j}|i\ra|m+j\ra.$$
By the definition of $m$,we have

{\small
$$d'-m\in\left\{
             \begin{array}{ll}
          \{1,2,...,d-1\} & d'\geq 2d \\
           \{1,2,...,r\}       & d<d'<2d, d'=d+r
             \end{array}
           \right.$$
}
hence in both case $d'-m<d$,but the Schmidt number of $v$ is no higher than $d'-m$.
So no  state in $V_2$ is maximally entangled.

From (i)(ii)(iii) and $dm<dd'$,we can conclude that the  states $\{|\phi_{i,j}\ra\}$ consist of a $dm$ member UMEB in
$\mathbb{C}^{d}\bigotimes\mathbb{C}^{d'}.$ \qed

\bigskip
From proposition 2 we can give a full answer to the question asked by \cite{s3,BC}.Compared with proposition 1,proposition
2 can solve the case when $d'=qd$.Moreover,it constructed different kinds of UMEB.

\bigskip
\noindent{\bf Example \ 2.}Now we give a 6 member UMEB in $\mathbb{C}^{2}\bigotimes\mathbb{C}^{4}$.(see Table II)

{\begin{table}[h]
 \caption{6-member UMEB in $\mathbb{C}^{2}\bigotimes\mathbb{C}^{4}$}

$\begin{array}{cccccccc}\hline\hline

  00 & 01 & 02 & 10 & 11 & 12 & 03 & 13 \\\hline
  1  &  0  & 0   & 0   & 1   & 0   & 0   &  0  \\
  1  &  0  &  0  &  0  & -1  & 0   & 0   &  0  \\
   0  &  1 &   0 &  0  &   0  & 1  & 0   & 0   \\
   0  &  1 &   0 &  0  &   0  & -1 & 0   &  0  \\
    0 &   0 & 1  & 1  &   0  &  0  & 0   &  0  \\
    0 &   0 &  1 & -1 &   0  &  0  & 0   &  0 \\ \hline\hline

\end{array}$
\end{table}

\bigskip
\noindent{\bf Example \ 3.}
Considering the UMEB in $\mathbb{C}^{3}\bigotimes\mathbb{C}^{6}$,$m$ can be chosen to be
4 or 5 by proposition 2,so there exist a 12 member and a 15 member UMEB.Now
we list the 15 member UMEB.(see Table III)

{\scriptsize
{\begin{table}[h]
 \caption{15-member UMEB in $\mathbb{C}^{3}\bigotimes\mathbb{C}^{6}$}
{\scriptsize
$
\begin{array}{cccccccccccccccccc}\hline\hline
  00 & 01 & 02 & 03 & 04 & 10 & 11 & 12 & 13 & 14 & 20 & 21 & 22 & 23 & 24 & 05 & 15 & 25 \\\hline
  1 & 0 & 0 & 0 & 0 & 0 & 1 & 0 & 0 & 0 & 0 & 0 & 1 & 0 & 0 & 0 & 0 & 0 \\
  1 & 0 & 0 & 0 & 0 & 0 & \omega & 0 & 0 & 0 & 0 & 0 & \omega^2 & 0 & 0 & 0 & 0 & 0 \\
  1 & 0 & 0 & 0 & 0 & 0 & \omega^2 & 0 & 0 & 0 & 0 & 0 & \omega & 0 & 0 & 0 & 0 & 0 \\
  0 & 1 & 0 & 0 & 0 & 0 & 0 & 1 & 0 & 0 & 0 & 0 & 0 & 1 & 0 & 0 & 0 & 0 \\
  0 & 1 & 0 & 0 & 0 & 0 & 0 & \omega & 0 & 0 & 0 & 0 & 0 & \omega^2 & 0 & 0 & 0 & 0 \\
  0 & 1 & 0 & 0 & 0 & 0 & 0 & \omega^2 & 0 & 0 & 0 & 0 & 0 & \omega & 0 & 0 & 0 & 0 \\
  0 & 0 & 1 & 0 & 0 & 0 & 0 & 0 & 1 & 0 & 0 & 0 & 0 & 0 & 1 & 0 & 0 & 0 \\
  0 & 0 & 1 & 0 & 0 & 0 & 0 & 0 & \omega & 0 & 0 & 0 & 0 & 0 & \omega^2 & 0 & 0 & 0 \\
  0 & 0 & 1 & 0 & 0 & 0 & 0 & 0 & \omega^2 & 0 & 0 & 0 & 0 & 0 & \omega & 0 & 0 & 0 \\
  0 & 0 & 0 & 1 & 0 & 0 & 0 & 0 & 0 & 1 & 1 & 0 & 0 & 0 & 0 & 0 & 0 & 0 \\
  0 & 0 & 0 & 1 & 0 & 0 & 0 & 0 & 0 & \omega & \omega^2 & 0 & 0 & 0 & 0 & 0 & 0 & 0 \\
  0 & 0 & 0 & 1 & 0 & 0 & 0 & 0 & 0 & \omega^2 & \omega & 0 & 0 & 0 & 0 & 0 & 0 & 0 \\
  0 & 0 & 0 & 0 & 1 & 1 & 0 & 0 & 0 & 0 & 0 & 1 & 0 & 0 & 0 & 0 & 0 & 0 \\
  0 & 0 & 0 & 0 & 1 & \omega & 0 & 0 & 0 & 0 & 0 & \omega^2 & 0 & 0 & 0 & 0 & 0 & 0 \\
  0 & 0 & 0 & 0 & 1 & \omega^2 & 0 & 0 & 0 & 0 & 0 & \omega & 0 & 0 & 0 & 0 & 0 & 0\\\hline\hline
\end{array}$
}
\end{table}

}
}

\section{conclusion}

We have studied the UMEB in $\mathbb{C}^{d}\bigotimes\mathbb{C}^{d'}(d\neq d')$.We conclude that there always
exist a UMEB in this case.Moreover,there might be two or more sets of UMEB with different numbers by proposition 2.
Meanwhile,we also give an answer to the question in \cite{BC} ,when $d=qd+r,q>1,0<r<d$ we extend them by some more $(q-1)d^2$ states to form a  $qd^2$ member UMEB.The main
difficult for proving a set to be UMEB is the unextendible condition.Different with the paper \cite{BC},we just use some
basic knowledge of the advance algebra to calculate the complementary space.

The result is just similar with the UPB problem in $\mathbb{C}^{d}\bigotimes\mathbb{C}^{d'}(d\neq d')$,
we can always find a UMEB.In the meantime,we also state that there are d-1 different sets of UMEB if $d'\geq2d$.

\bigskip

$Acknowledgments:$ We thank  professor Shao-Ming Fei from Capital Normal University provided us this question
and gave us some helpful advices.

\end{document}